\documentclass[twocolumn,showkeys,aps,prb,showpacs]{revtex4-1}
\usepackage{graphicx}
\usepackage{amsmath}
\usepackage[CJKbookmarks,dvipdfm,colorlinks,linkcolor=blue,citecolor=blue]{hyperref}

\begin{document}

\title{Ultrahigh  lattice thermal conductivity  in topological semimetal TaN caused by large acoustic-optical gap}
\author{San-Dong Guo$^1$}
\author{Bang-Gui Liu$^{2,3}$}
\affiliation{$^1$School of Physics, China University of Mining and
Technology, Xuzhou 221116, Jiangsu, China}
\affiliation{$^2$Beijing National Laboratory for Condensed Matter Physics,
Institute of Physics, Chinese Academy of Sciences, Beijing 100190, China}
\affiliation{$^3$School of Physical Sciences, University of Chinese Academy of Sciences, Beijing 100190, China}

\begin{abstract}
Topological semimetal  may have potential applications like topological qubits, spintronics and quantum computations. Efficient heat dissipation is a key factor for the reliability and stability of topological semimetal-based nano-electronics devices, which is closely related to high thermal conductivity.
 In this work, the elastic properties and lattice thermal conductivity  of  TaN are investigated  by  first-principles calculations and the linearized phonon Boltzmann equation within the single-mode relaxation time approximation (RTA). According to the calculated bulk modulus, shear modulus and $C_{44}$, TaN can be regarded as a potential  incompressible and hard material.
 The  room-temperature lattice thermal conductivity  is predicted to be 838.62   $\mathrm{W m^{-1} K^{-1}}$ along a  axis and  1080.40   $\mathrm{W m^{-1} K^{-1}}$ along  c axis, showing  very strong anisotropy. It is found that the lattice thermal conductivity of TaN is  several tens of times higher than one of other topological semimetal, such as TaAs, MoP and ZrTe, which is due to very longer phonon lifetimes for TaN than other topological semimetal.  The very different atomic masses of Ta and N atoms lead to a very large acoustic-optical band gap, and then prohibits the scattering between acoustic and optical phonon modes,  which gives rise to very long phonon lifetimes. Based on mass difference factor, the WC and WN can be regarded as potential candidates with ultrahigh  lattice thermal conductivity. Calculated results show that isotope scattering  has  little effect on lattice thermal conductivity, and that phonon with mean free path(MFP) larger than 20 (80) $\mathrm{\mu m}$ at 300 K  has little contribution  to the total lattice thermal conductivity.
 This work implies that  TaN-based nano-electronics devices may be more stable and reliable due to efficient heat dissipation, and motivate  further experimental works to study lattice thermal conductivity of TaN.

\end{abstract}

\pacs{72.15.Jf, 71.20.-b, 71.70.Ej, 79.10.-n ~~~~~~~~~~~~~~~~~~~~~~~~~~~~~~~~~~~Email:sandongyuwang@163.com}

\maketitle

\section{Introduction}
Topological nontrivial phase, including topological insulator and semimetal,  is one of the major advancements
in condensed matter physics and material science\cite{q6,q7,q8,q1,q3,q4,q2,q10,q10-3,q10-4,q5,q5-1,q10-1,q10-2,n1}.
The representative topological semimetals, such as  Dirac semimetal ($\mathrm{Na_3Bi}$)\cite{q4}, Weyl semimetal (TaAs)\cite{q10,q10-1,q10-2,q10-3} and   nodal line semimetal (ZrSiS)\cite{n1},  have
been confirmed by angle-resolved photoemission spectroscopy (ARPES).
In Dirac   and Weyl semimetals,  even-fold degenerate  point can be observed in the momentum space\cite{q4,q10,q5,q5-1}, such as four-fold degenerate Dirac point and two-fold degenerate Weyl point.
Beyond Dirac and Weyl fermions, three-, six- or eight-fold band crossings are proposed as new types of topological semimetals\cite{j1}.
A band crossing between a
doubly degenerate band and a nondegenerate band, namely three-fold degenerate crossing points, has been predicted in TaN,  MoP and ZrTe with  WC-type structure \cite{q11-00,q11,q11-0}, and in in $\mathrm{InAs_{0.5}Sb_{0.5}}$\cite{q11+0}. Experimentally, the MoP has been confirmed to be topological semimetal with triply degenerate nodal points (TDNPs), coexisting with the pairs of Weyl points\cite{q7}.
\begin{figure}
  \includegraphics[width=7cm]{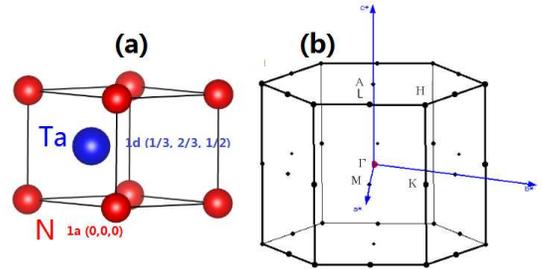}
  \caption{(Color online)(a)The crystal structure of TaN in one unit cell; (b)the  Brillouin zone with high-symmetry points.}\label{st}
\end{figure}

These topological semimetals may have  substantial applications in electronics, spintronics and quantum computation\cite{k1}. Efficient heat dissipation is a key factor for the reliable performance and stable function of electronic devices
based on these topological semimetals, especially for the high-power
situations\cite{k2}. High lattice thermal conductivity is in favour of the high-performance thermal management\cite{k3}.
The lattice thermal conductivity in TaAs, MoP and ZrTe has been calculated from a first principles calculation, showing  obvious anisotropy
along the a and c crystal axis\cite{gsd1,gsd2,q12,q13}.  However, their lattice thermal conductivities are relatively low, about 17$\sim$44 $\mathrm{W m^{-1} K^{-1}}$ at 300 K, which  is against  efficient heat dissipation. Therefore, searching for topological semimetals with high  lattice thermal conductivity  is very necessary and interesting. In this work, the elastic properties of  topological semimetal TaN are studied from first-principles calculations, and the phonon transport properties  are performed by solving the phonon Boltzmann transport equation. The
calculated  bulk modulus, shear modulus and $C_{44}$ suggest that  TaN is a potential low compressible and hard material.
The calculated lattice thermal conductivity is very higher than one of  TaAs, MoP and ZrTe\cite{gsd1,gsd2,q12,q13}, and the room-temperature lattice thermal conductivity  is predicted to be 838.62   $\mathrm{W m^{-1} K^{-1}}$ and  1080.40   $\mathrm{W m^{-1} K^{-1}}$ along the a and c axis. This can be attributed to the
large acoustic-optical frequency gap due to the large mass difference
of Ta and N, producing  inefficient scattering among
acoustic and optical phonon modes. The
mass difference factor suggests  that
WC and WN are potential topological  materials with ultrahigh  lattice thermal conductivity.

The rest of the paper is organized as follows. In the next
section, we shall give our computational details about phonon transport. In the third section, we shall present elastic and  phonon transport properties of TaN. Finally, we shall give our discussion and  conclusions in the fourth section.

\begin{figure}
  \includegraphics[width=8cm]{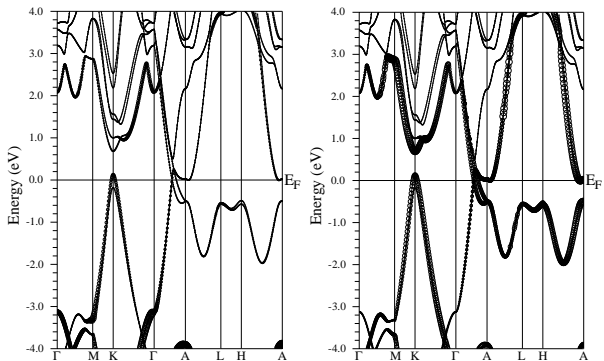}
  \caption{The calculated energy  band structures of TaN along high-symmetry paths using GGA+SOC, with fatted bands projected onto N (left) and Ta (Right) atoms  .}\label{band}
\end{figure}

\section{Computational detail}
First-principles calculations are carried out within the projected augmented wave (PAW) method, as implemented in the VASP code\cite{pv1,pv2,pv3}.
The generalized gradient approximation of the Perdew-Burke-Ernzerhof (GGA-PBE)\cite{pbe} is adopted  for the exchange-correlation
functional with the plane-wave-cut-off energy of 500 eV,
The 2s and 2p electrons of N,  and  6s and 5d electrons of Ta  are treated as valance ones.
The energy convergences are less than  $10^{-8}$ eV.
The  lattice thermal conductivity of  TaN  is calculated by solving linearized phonon Boltzmann equation with the single mode RTA,   as implemented in the Phono3py code\cite{pv4}. The lattice thermal conductivity can be expressed as
\begin{equation}\label{eq0}
    \kappa_L=\frac{1}{NV_0}\sum_\lambda \kappa_\lambda=\frac{1}{NV_0}\sum_\lambda C_\lambda \nu_\lambda \otimes \nu_\lambda \tau_\lambda
\end{equation}
where $\lambda$ is phonon mode, $N$ is the total number of q points sampling the  Brillouin zone (BZ), $V_0$ is the volume of a unit cell, and  $C_\lambda$,  $ \nu_\lambda$, $\tau_\lambda$   is the specific heat,  phonon velocity,  phonon lifetime.
The phonon lifetime $\tau_\lambda$ can be attained by  phonon linewidth $2\Gamma_\lambda(\omega_\lambda)$ of the phonon mode
$\lambda$:
\begin{equation}\label{eq0}
    \tau_\lambda=\frac{1}{2\Gamma_\lambda(\omega_\lambda)}
\end{equation}
The $\Gamma_\lambda(\omega)$  takes the form analogous to the Fermi golden rule:
\begin{equation}
\begin{split}
   \Gamma_\lambda(\omega)=\frac{18\pi}{\hbar^2}\sum_{\lambda^{'}\lambda^{''}}|\Phi_{-\lambda\lambda^{'}\lambda^{''}}|^2
   [(f_\lambda^{'}+f_\lambda^{''}+1)\delta(\omega
    -\omega_\lambda^{'}-\\\omega_\lambda^{''})
   +(f_\lambda^{'}-f_\lambda^{''})[\delta(\omega
    +\omega_\lambda^{'}-\omega_\lambda^{''})-\delta(\omega
    -\omega_\lambda^{'}+\omega_\lambda^{''})]]
\end{split}
\end{equation}
in which $f_\lambda$ is the phonon equilibrium occupancy and
$\Phi_{-\lambda\lambda^{'}\lambda^{''}}$
is the strength of interaction among the three phonons $\lambda$, $\lambda^{'}$,
and $\lambda^{''}$ involved in the scattering.

The second- and third-order interatomic force constants (IFCs) are calculated
by the supercell approach  with finite atomic displacement
of 0.03 $\mathrm{{\AA}}$ .
For second-order harmonic IFCs, a 4 $\times$ 4 $\times$ 4  supercell  containing
128 atoms is used  with k-point meshes of 2 $\times$ 2 $\times$ 2. Based on the harmonic IFCs, phonon dispersion of TaN can be calculated by  Phonopy package\cite{pv5}, which  determines the allowed three-phonon scattering processes. The  group velocity  and specific heat can also  be attained from phonon dispersion. For the third-order anharmonic IFCs,  a 3 $\times$ 3 $\times$ 3
supercells containing 54 atoms is used  with k-point meshes of 3 $\times$ 3 $\times$ 3.
 Based on third-order anharmonic IFCs, the three-phonon scattering rate can be attained, and  further   the phonon lifetimes can be calculated. To compute lattice thermal conductivities, the reciprocal spaces of the primitive cells  are sampled using the 20 $\times$ 20 $\times$ 20 meshes.
\begin{figure}
  \includegraphics[width=8cm]{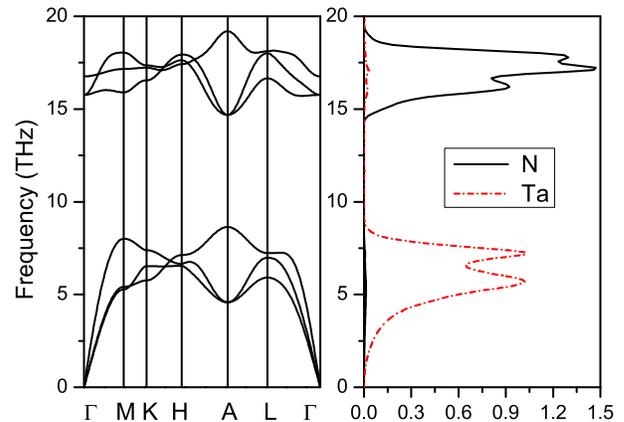}
  \caption{(Color online)Phonon  dispersion curve of TaN, along  with atom partial density of states (PDOS). }\label{ph}
\end{figure}

\begin{table*}
\centering \caption{The elastic constants $C_{ij}$, bulk ($B$), shear ($G$) and Young's ($E_{xx}$ and $E_{zz}$) moduli  (in GPa) of TaN, MoP and ZrTe. }\label{tab}
  \begin{tabular*}{0.96\textwidth}{@{\extracolsep{\fill}}ccccccccccc}
  \hline\hline
 Name & $C_{11}$ & $C_{12}$& $C_{13}$&$C_{33}$ &$C_{44}$&$C_{66}$&$B$&$G$&$E_{xx}$&$E_{zz}$\\\hline\hline
TaN&566.40&128.23&62.41&706.39  &215.02&219.09&260.26&233.16&534.04&695.17\\\hline
MoP&359.00&153.73&160.14&515.15  &169.22&102.64&239.10&134.96&274.28&415.11\\\hline
ZrTe&140.82&58.78&88.81&201.11  &110.36&  41.02&102.50&61.26&97.83&122.07\\\hline\hline
\end{tabular*}
\end{table*}
\begin{figure}
  \includegraphics[width=8cm]{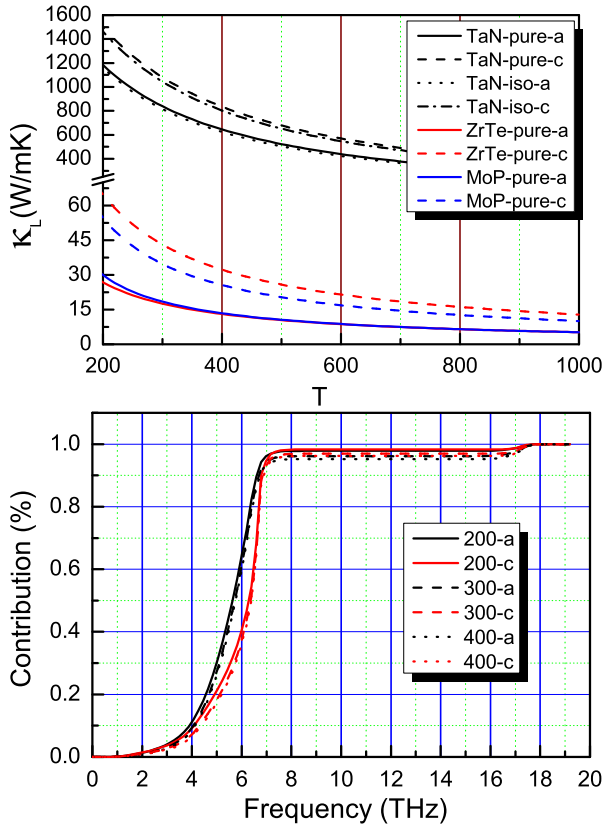}
  \caption{(Color online) The lattice thermal conductivities  of infinite (Pure and Isotope) TaN, ZrTe and MoP as a function of temperature, including a  and c directions; The cumulative lattice thermal conductivity (200, 300 and 400 K) of  infinite (Pure) TaN  divided by total lattice thermal conductivity with respect to phonon frequency, along a and c directions.}\label{kl}
\end{figure}

\section{MAIN CALCULATED RESULTS AND ANALYSIS}
The  TaN shows  WC-type crystal structure with space group  $P\bar{6}m2$ (No.187),  and the crystal structure is  shown in \autoref{st}. The MoP and ZrTe have the same crystal structure with TaN. The Ta and
N atoms occupy the 1d (1/3, 2/3, 1/2) and 1a (0,0,0) Wyckoff positions, respectively. Experimentally, it can be synthesized at
high pressure  within a proper high temperature range.
The experimental lattice constants ($a$=$b$=2.9333 $\mathrm{{\AA}}$, $c$=2.8844 $\mathrm{{\AA}}$ )\cite{q14} are used to investigate elastic properties and  lattice thermal conductivity  of TaN.
Firstly, the energy band structures of TaN with fatted bands projected onto N  and Ta  atoms  are  plotted in \autoref{band}. It can be seen that there are two TDNPs along $\Gamma$-A direction with Ta-d character. Our calculated energy band structures agree well with previous report\cite{q11-00}.
For MoP and ZrTe,  besides TDNPs, there are Weyl nodes in first BZ\cite{q7,q11}.

Based on the experimental crystal structure,
the elastic constants $C_{ij}$ of the TaN are calculated, which are listed in \autoref{tab}, along with ones of MoP and ZrTe. The elastic constants
satisfy the following mechanical stability criteria\cite{el,q15}£º
\begin{equation}\label{e1}
C_{44}>0
\end{equation}
\begin{equation}\label{e1}
 C_{11}>|C_{12}|
\end{equation}
\begin{equation}\label{e1}
(C_{11}+2C_{12})C_{33}>2C_{13}^2
\end{equation}
This indicates   that the system of TaN is in a mechanical stable.
Based on calculated elastic constants, the bulk, shear and Young's modulus can be obtained by Voigt-Reuss-Hill
approximations. The Voigt's, Reuss's and Hill's bulk modulus can be calculated  by the following equations:
\begin{equation}\label{4}
    B_V=\frac{1}{9}(2C_{11}+C_{33}+2C_{12}+4C_{13})
\end{equation}
\begin{equation}\label{5}
    B_R=(2S_{11}+S_{33}+2S_{12}+4S_{13})^{-1}
\end{equation}
\begin{equation}\label{6}
    B_H=\frac{1}{2}(B_V+B_R)
\end{equation}
The Voigt's, Reuss's and Hill's shear modulus can be attained by using these formulas:
\begin{equation}\label{4}
    G_V=\frac{1}{15}(2C_{11}+C_{33}-C_{12}-2C_{13}+6C_{44}+3C_{66})
\end{equation}
\begin{equation}\label{5}
    G_R=[\frac{1}{15}(8S_{11}+4S_{33}-4S_{12}-8S_{13}+6S_{44}+3S_{66})]^{-1}
\end{equation}
\begin{equation}\label{6}
    G_H=\frac{1}{2}(G_V+G_R)
\end{equation}
The Young's modulus  $E_{ii}$  can be computed  by the relationship:
\begin{equation}\label{23}
    E_{ii}=1/S_{ii}
\end{equation}
The $S_{ij}$ are the elastic compliance constants.
The related data are tabulated in \autoref{tab}, together with ones of MoP and ZrTe.
The $B$/$G$ can be used to measure material behaviour as ductile ($B$/$G$$>$1.75) or brittle ($B$/$G$$<$1.75). For TaN, the calculated $B/G$ ratio value  is 1.12, indicating that the brittle character is dominant.  Bulk modulus or shear modulus
can measure the hardness of materials\cite{h1}.  The materials with higher  bulk or shear modulus may be  likely  harder
materials. The magnitude of the shear modulus $C_{44}$ may be a better hardness
predictor for transition-metal carbonitrides\cite{h2,h3}. Based on  these criteria, TaN may be a incompressible and hard material.
 Experimental studies on structural and mechanical properties of TaN are strongly recommended.
\begin{figure}
  \includegraphics[width=8cm]{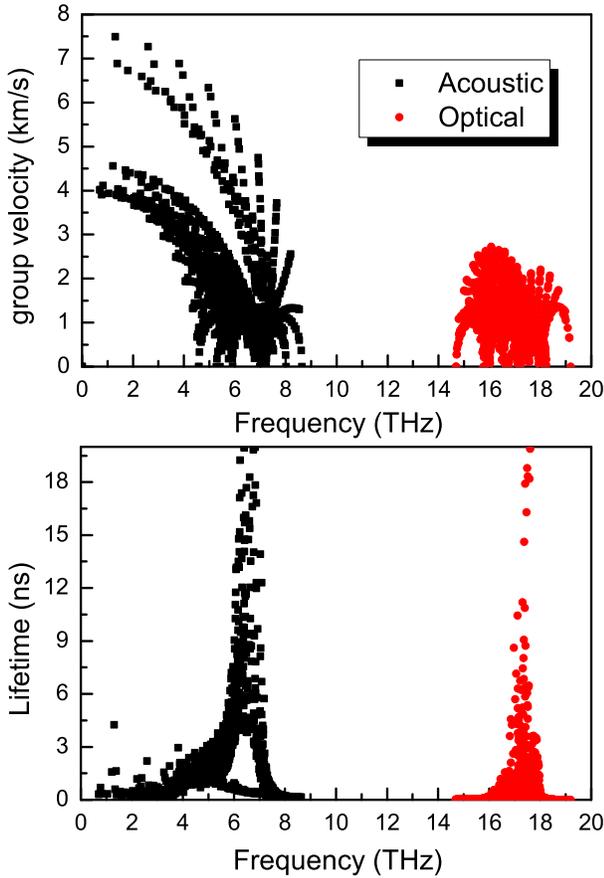}
  \caption{(Color online)The mode level phonon group velocities and  phonon lifetimes (300K) of infinite (Pure) TaN in the first BZ.}\label{v}
\end{figure}

\begin{figure}
  \includegraphics[width=8cm]{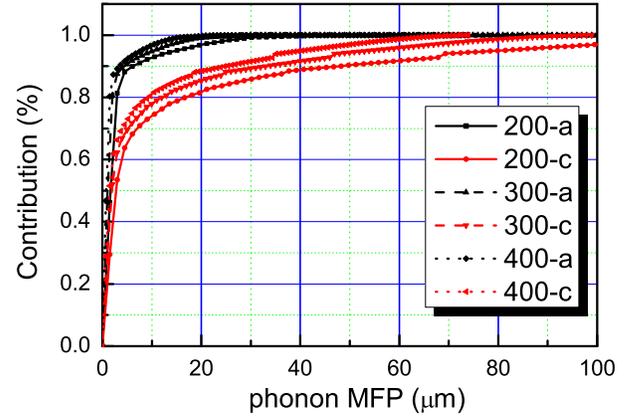}
  \caption{(Color online)At 200, 300 and 400 K, the cumulative lattice thermal conductivity divided by total lattice thermal conductivity  with respect to phonon MFP along a and c directions.}\label{mfp}
 \end{figure}6

Based on harmonic IFCs matrix, the phonon dispersion and and atom partial density of states (DOS) of
 TaN  are plotted in \autoref{ph}.  No imaginary frequencies are observed in the phonon dispersion of
TaN,  indicating  the thermodynamic stability of TaN.
Two atoms  per unit cell lead to 3 acoustic and 3 optical phonon branches.
In contrast
to MoP and ZrTe,  a significant difference is observed.
It is clearly seen  that
there is  a  phonon band gap of 6.05 THz  at the A point (0, 0, $\pi$/2) between  acoustic and optical branches, which is very larger than one of MoP (0.68 THz) and ZrTe (0.15 THz)\cite{gsd1,gsd2}.
 The large  phonon band gap is  due to very different atomic masses of Ta and N atoms\cite{m1,m3}.
  A mass difference factor,  defined as $\delta=(M_{max}-M_{min})/M_{min}$, is used to measure the different  strength, and the corresponding value is 11.92 for TaN, 2.10 for MoP and  0.40 for ZrTe.
According to atom partial DOS (PDOS),  the vibrations of
Ta (N) atoms almost completely dominate  acoustic (optical) branches. These are  familiar from
the diatomic linear chain model,  where acoustic (optical) branches
are mainly contributed by the vibrations of by the larger (smaller) mass.

From harmonic and anharmonic IFCs, the intrinsic lattice thermal conductivity of TaN  can be calculated by solving the linearized phonon Boltzmann equation within single-mode RTA method.  Based on the formula proposed  by Shin-ichiro Tamura\cite{q24}, the phonon-isotope scattering can be included.
Along a and c directions, the lattice thermal conductivities  of pure and   isotopic TaN  as a function of temperature are shown in \autoref{kl}, together with ones of ZrTe and MoP for comparison.  Similar to ZrTe and MoP,
 the lattice thermal conductivity of TaN
 shows obvious anisotropy.   It is clearly seen that   the  c-axis lattice thermal conductivity   is very  higher than a-axis one.
The room-temperature  lattice thermal conductivity  of pure (isotopic) TaN  along a and c axis  is 838.62 (814.96)  $\mathrm{W m^{-1} K^{-1}}$ and  1080.40 (1044.06)  $\mathrm{W m^{-1} K^{-1}}$, respectively. The room temperature
"isotope effect" is given by $P=(\kappa_{pure}/\kappa_{iso}-1)$, which is 2.90\% along a axis, and 3.48\% along c axis.
These mean that phonon-isotope scattering has little effects on lattice thermal conductivity. Due to enhancement of phonon-phonon scattering, isotopic effect on lattice thermal conductivity gradually becomes weak with increasing temperature.
 An anisotropy factor\cite{q12}, defined as $\eta=(\kappa_{L}(cc)-\kappa_{L}(aa))/\kappa_{L}(aa)$, can be used to  measure the anisotropic strength.
 The $\eta$ for TaN is 28.83\%, which is smaller than that of MoP (88.5\%) and ZrTe (145.3\%), implying weak anisotropy with respect to MoP  and ZrTe.
This can also be explained by the shear
anisotropy ratio $A$, defined as  $A=C_{44}/C_{66}$\cite{h4}. The calculated $A$  for TaN (0.98) is more closer to 1 than ones of MoP (1.65) and ZrTe (2.69).
  It is noted that the lattice thermal conductivity of TaN along a (c) direction is around  45  (31) times higher than that of MoP, and about 48 (25) times higher than one of ZrTe.
 The relation between lattice thermal conductivity and Young's modulus is $\kappa_L\sim \sqrt{E}$\cite{q16}. It is found that the order of lattice thermal conductivity along a and c directions is consistent with one of Young's modulus.
 At 200, 300 and 400 K, the cumulative lattice thermal conductivities divided by total lattice thermal conductivity with respect to frequency  along a and c directions are shown in \autoref{kl}. The cumulative thermal conductivity is defined by:
\begin{equation}\label{eq2}
  \kappa^c(\omega)=\int^\omega_0\Sigma_\lambda \kappa_\lambda\delta(\omega_\lambda-\omega^{'}) d \omega^{'}
\end{equation}
It is clearly seen  that nearly all lattice thermal conductivity is made up of  the acoustic phonon branches for three considered temperature.  It is noted that the cumulative lattice thermal conductivities divided by total lattice thermal conductivity with respect to frequency has weak  temperature dependence. The slope along a direction is larger than that along c direction, which means that low frequency phonon has larger contribution to total lattice thermal conductivity
for a than c direction.

\begin{figure}
  \includegraphics[width=8cm]{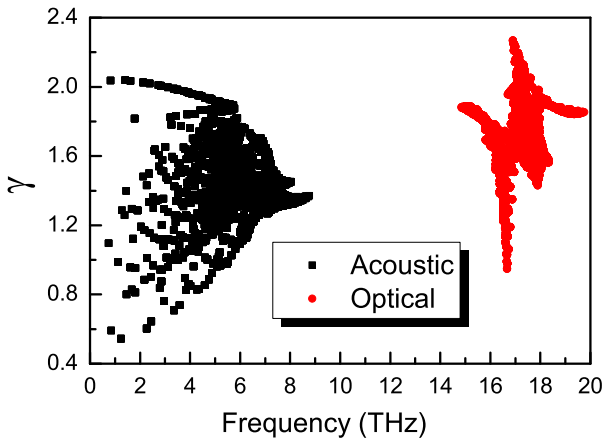}
  \caption{(Color online)The mode level phonon  Gr$\mathrm{\ddot{u}}$neisen parameters ($\gamma$) of TaN in the first BZ.}\label{r}
\end{figure}

To gain more insight into high  lattice thermal conductivity of TaN,  we show  the mode level phonon group velocities
and lifetimes  in \autoref{v}. In long-wavelength limit, the largest phonon group velocity  of  TA1, TA2 and LA branches  is 3.91 $\mathrm{km s^{-1}}$, 4.12 $\mathrm{km s^{-1}}$ and  6.89 $\mathrm{km s^{-1}}$, respectively. These are larger than ones of ZrTe\cite{gsd2}, but smaller than ones of MoP\cite{gsd1}.
It  is clearly seen  that  the most of group
velocities of  acoustic branches are higher than those of optical branches.
It is found that most of   phonon lifetimes of TaN are very longer than ones of MoP and ZrTe\cite{gsd1,gsd2}, which leads to very higher lattice thermal conductivity for TaN than MoP and ZrTe. Unexpectedly,  the phonon lifetimes  near 6 and 17 THz become very large. This can be understood by very large frequency gap between the optical and acoustic phonon branches. The frequency gap (6.05 THz)  is close to
the range of acoustic phonons (8.64 THz). While heat is transmitted primarily by the
 acoustic branches, the optical branches provide
important scattering channels for the acoustic modes, particularly through acoustic+acoustic$\rightarrow$optical scattering.
Because of  the requirement on energy conservation
for phonon-phonon scattering,  the annihilation process of two acoustic phonon
modes into one optical one becomes ineffective (such annihilation process
is not totally prohibited.) caused by acoustic-optical gap. As a result, the weaker phonon-phonon scattering
rate is produced, and then results in long phonon lifetimes,  giving rise to a much high lattice thermal
conductivity. The very high thermal conductivity  is found in  BAs\cite{h5} and AlSb\cite{h6}, which is also  due to a large frequency gap.

\begin{table*}
\centering \caption{For TaN, MoP and ZrTe, the longitudinal, transversal  and average sound speed ($v_L$, $ v_T$ and $v_A$ in km/s), Debye temperature ($T_D$ in K); the frequency gap between acoustic and optical phonons ($Gap_{ao}$ in THz); the a-axis, c-axis and average lattice thermal conductivity ($\kappa_{aa}$, $\kappa_{cc}$ and $\kappa_A$ in $\mathrm{W m^{-1} K^{-1}}$).}\label{tab1}
  \begin{tabular*}{0.96\textwidth}{@{\extracolsep{\fill}}ccccccccc}
  \hline\hline
 Name & $v_L$ &$ v_T$& $v_A$&$T_D$ &$Gap_{ao}$&$\kappa_{aa}$&$\kappa_{cc}$&$\kappa_A$\\\hline\hline
TaN&6.11&3.79&4.80&647.92 &6.05&838.62&1080.4&919.21\\\hline
MoP&7.48&4.36&5.63&687.17 &0.68&18.41&34.71&23.84\\\hline
ZrTe&4.92&3.12&3.92&405.78  &0.15&  17.56&43.08&26.07\\\hline\hline
\end{tabular*}
\end{table*}

The size dependence of lattice thermal
conductivity of TaN can be reflected by the cumulative lattice thermal conductivity  with respect to MFP, which shows
 how phonons with different MFP contribute to the thermal conductivity.
The MFP cumulative lattice thermal conductivity is defined as:
\begin{equation}\label{eq2}
  \kappa^c(l)=\int^l_0\Sigma_\lambda \kappa_\lambda\delta(l_\lambda-l^{'}) d l^{'}
\end{equation}
\begin{equation}\label{eq2}
  l_\lambda=|\mathrm{l_\lambda}|=|\mathrm{\nu_\lambda\otimes\tau_\lambda}|
\end{equation}
At 200, 300 and 400 K, the cumulative lattice thermal conductivity divided by total lattice thermal conductivity  with respect to MFP   along a and c directions are  shown in \autoref{mfp}.
 It is clearly seen that the cumulative lattice thermal conductivity of TaN along both a and c axis  approaches saturation value  with MFP increasing.
 With the increasing temperature, the critical MFP gradually decreases.  At 300 K,  phonons with MFP larger than 20 (84) $\mathrm{\mu m}$  along a (c) direction has little contribution  to the total lattice thermal conductivity.
At room temperature, phonons with MFP smaller than 1 (2) $\mathrm{\mu m}$ along a(c) direction
 contribute  around half  to the total lattice thermal conductivity. These results mean that the lattice thermal conductivity along c direction is tuned more easily than that along a direction.

\section{Discussions and Conclusion}
\begin{figure}
  \includegraphics[width=8cm]{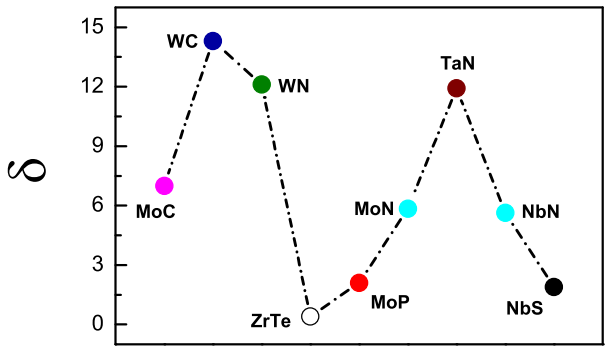}
  \caption{(Color online)The mass difference factor of  a list of  topological metals with  WC-type crystal structure.}\label{md}
\end{figure}

The MoP, ZrTe and TaN have the same crystal structure, and their lattice thermal conductivities show  obvious anisotropy along a and c directions, where the lattice thermal conductivity along c direction is larger than that along a direction. However, a very higher lattice thermal conductivity of TaN is observed than that of MoP and ZrTe. Traditionally, a low Debye temperature $T_D$ indicates low lattice
thermal conductivity.  All
phonon modes are excited with the temperature  above $T_D$, which can give rise to strong three-phonon
scattering, and then  suppress lattice thermal conductivity. When the temperature is below
$T_D$, some phonon modes begin to be frozen out\cite{h7}.
The Debye temperature  can be
obtained from the average sound velocity using the following equation\cite{h8}:
\begin{equation}\label{eq2}
 T_D=\left(\frac{3N}{4\pi V_0}\right)^{1/3}\frac{hv_{ A}}{k_{B}}
\end{equation}
where N denotes the number of atom in the primitive unit cell, $V_0$ denotes the unit cell volume, $h$ and $k_{B}$ denote the Planck's and Boltzmann's constants. The average sound speed $v_A$ can be calculated from  the longitudinal and transversal sound velocities, $v_L$ and $v_T$.
\begin{equation}\label{eq2}
 v_{A}=\left[\frac{1}{3}\left(\frac{1}{v_{L}^3}+\frac{2}{v_{T}^3}\right)\right]^{-1/3}
\end{equation}
The longitudinal and transversal sound velocities, $v_L$ and $v_T$, are related to the bulk, shear  modulus and the density of the material, $B$, $G$ and $\rho$.
\begin{equation}\label{eq2}
v_{L}=\sqrt{(B+4G/3)/\rho}
\end{equation}
\begin{equation}\label{eq2}
v_{T}=\sqrt{G/\rho}
\end{equation}
The longitudinal, transversal, average sound speed  and  Debye temperature of TaN, MoP and ZrTe are listed in \autoref{tab1}.
Assumed from Debye temperature, the lattice thermal conductivity of TaN should be lower than one of MoP, and
  should be slightly higher than that of ZrTe. In contrast to this straightforward prediction,
the calculated lattice thermal conductivity of TaN is
dozens of times higher than one of MoP or ZrTe. Based on the formula proposed  by Slack\cite{h9}, four factors, including the  average atomic mass,  interatomic bonding,
crystal structure and  anharmonicity, determine the lattice thermal conductivity.  A high
Debye temperature can be produced by low average
atomic mass and strong interatomic bonding, leading to a high thermal conductivity.
\begin{figure}
  \includegraphics[width=8cm]{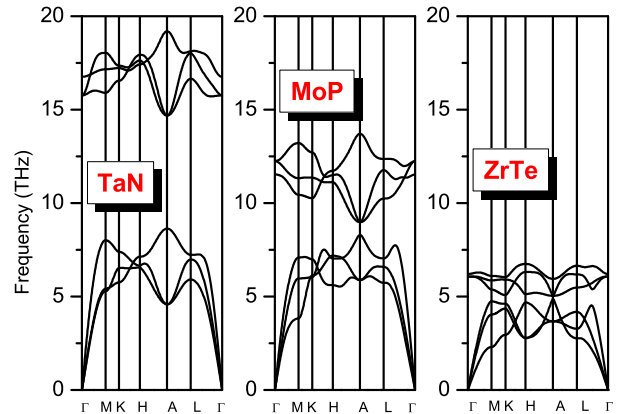}
  \caption{(Color online)Phonon  dispersion curves of TaN, MoP and ZrTe. }\label{ph-1}
\end{figure}
Mode Gr$\mathrm{\ddot{u}}$neisen parameters $\gamma$ can measure the strength of anharmonic interactions, determining the intrinsic phonon-phonon scattering. The larger  $\gamma$ gives rise  to stronger anharmonic phonon scattering, leading to lower  lattice thermal conductivity.
 The mode level  Gr$\mathrm{\ddot{u}}$neisen parameters of  TaN are plotted in \autoref{r}.  The average  Gr$\mathrm{\ddot{u}}$neisen parameter is 1.60, which is close to 1.57 of MoP\cite{gsd1} and 1.52 of ZrTe\cite{gsd2}.  Therefore, other factors should be found to explain very high lattice thermal conductivity in TaN.
 Although TaN, MoP and ZrTe have similar outlines of phonon dispersion (see \autoref{ph-1}),  a significant difference is found. A very large frequency gap (6.05 THz) between the optical and acoustic phonon branches in TaN  is observed,  while the
gap is only  0.68 THz for MoP  and 0.15 THz for ZrTe. The large thermal conductivity of TaN is due  to the
large acoustic-optical frequency gap caused by  the large mass difference
of Ta and N, which  can  lead to inefficient scattering among acoustic and optical phonon modes.
The frequency gap between acoustic and optical phonons,   a-axis, c-axis and  average  lattice thermal conductivities ($\kappa_A$=($\kappa_{aa}$+$\kappa_{bb}$+$\kappa_{cc}$/3) of TaN, MoP and ZrTe  are listed in \autoref{tab1}. Therefore, the
extremely large frequency gap has a dramatic effect
on lattice thermal conductivity of materials. Recently,  many materials with  WC-type crystal structure have been predicted  as topological metal
candidates\cite{q4}.  The mass difference factor $\delta$ for these  topological metals  are plotted in \autoref{md}.  The large $\delta$ can lead to large acoustic-optical gap, inducing large lattice thermal conductivity by restricting acoustic+acoustic$\rightarrow$optical scattering. It is clearly seen that $\delta$ of WC and WN is larger than 10, being close to one of TaN. So, WC and WN may be potential candidates with ultrahigh  lattice thermal conductivity.

In summary,  the elastic and phonon  transport properties of TaN are investigated  by   combining the first-principles calculations and semiclassical Boltzmann transport theory. Based on elastic tensor components $C_{ij}$, the mechanical stability of TaN is confirmed by  mechanical stability criteria.
 The bulk modulus, shear modulus, Young's modulus, the longitudinal  sound speed, transversal  sound speed and  Debye temperature are also attained, according to calculated  $C_{ij}$. It is predicted  that
TaN may be a potential low compressible and hard material,  based on calculated bulk modulus, shear modulus, and $C_{44}$.
It is found that TaN has
ultrahigh  lattice thermal conductivity, showing  an obvious anisotropy along the a and c crystal axis. The
extremely large frequency gap in TaN strongly restricts acoustic+acoustic$\rightarrow$optical scattering through
the energy conservation, which leads to ultrahigh  lattice thermal conductivity in TaN.
Calculated results show that isotope
scattering has  little effect on the lattice thermal conductivity of TaN, and  phonons with MFP larger than 20 (80) $\mathrm{\mu m}$  along a (c) direction have little contribution to the total lattice thermal conductivity.
Our works shed light on elastic and phonon transport properties of TaN, and help to seek ultrahigh  lattice thermal conductivity in topological semimetals by large frequency gap, such as WC and WN.
\begin{acknowledgments}
This work is supported by the National Natural Science Foundation of China (Grant No.11404391). We are grateful to the Advanced Analysis and Computation Center of CUMT for the award of CPU hours to accomplish this work.
\end{acknowledgments}

\end{document}